%% file: main.tex
\documentclass[a4paper,11pt]{article}
\pdfoutput=1 

\usepackage{jinstpub} 
\usepackage{xcolor}
\usepackage[colorinlistoftodos,prependcaption,textsize=tiny]{todonotes} 
\usepackage{lineno}
\usepackage{siunitx}
\title{\boldmath High voltage scheme optimization for secondary discharge mitigation in GEM-based detectors}

\author[a]{L. Lautner}
\author[a,b]{L. Fabbietti}
\author[a,b]{P. Gasik}
\author[a]{T. Klemenz}
\affiliation[a]{Physik Department E62, Technische Universit\"{a}t M\"{u}unchen, James-Franck-Str. 1, 85748 Garching, Germany}
\affiliation[b]{Excellence Cluster Universe, Technische Universit\"{a}t M\"{u}nchen, Boltzmannstr. 2, 85748 Garching, Germany}

\emailAdd{lukas.lautner@tum.de}
\abstract{We investigate the influence of the high voltage scheme elements on the stability of a detector based on a single $10\times10$\,cm$^2$ area GEM with respect to the secondary discharge occurrence. These violent events pose a major threat to the integrity of GEM detectors and their Front-End Electronics and need to be avoided by any means. For a single GEM setup, we propose a detailed high voltage scheme that is designed to prevent secondary discharges. We determine optimal values of the protection resistors and parasitic capacitances introduced by cables used in the system. The results of this paper may be used as a guideline for the optimization of more complicated multi-GEM detectors.}
\keywords{Electron multipliers (gas), Micropattern gaseous detectors, Spark discharge, Secondary discharges, Discharge propagation}
\arxivnumber{} 

\begin{document}
\maketitle
\raggedbottom

\input{Introduction}
\input{Setup}

\input{SingleGEM}
\input{Conclusion}


\acknowledgments


The Authors wish to thank the ALICE TPC Collaboration (in particular H. Appelsh\"{a}user, C. Garabatos, R. M\"{u}nzer, A. Deisting) and the RD51 Collaboration (in particular F. Sauli, L. Ropelewski, E. Oliveri, H. M\"{u}ller) for fruitful discussions.

This work was supported by the Deutsche Forschungsgemeinschaft (DFG, German Research Foundation) [grant number FA 898/4-1] and by the DFG Cluster of Excellence "Origin and Structure of the Universe" (www.universe-cluster.de) [project number DFG 492 EXC153].


\end{document}

%% file: Introduction.tex
\section{Introduction}
\label{sec:introduction}

Detectors based on Gas Electron Multipliers (GEMs)~\cite{Sauli} are employed in many high-rate collider and fixed-target experiments such as COMPASS~\cite{compass_Ketzer,compass_Altunbas}, LHCb~\cite{LHCb_Bencivenni} or TOTEM~\cite{totem_Bozzo,totem_Bagliesi}. GEMs are baseline technology of the ongoing upgrades of the CMS Moun Endcap~\cite{cms_Colaleo} and the ALICE TPC~\cite{alice_Ketzer,alice_Lippmann}, and planned apparatus, like the
sPHENIX TPC~\cite{sphenix_Aidala} . The long-term operation of such detectors in the harsh environment of high-rate experiments puts high requirements on radiation hardness, ageing resistance, and stability against electrical discharges. Especially the latter pose a threat to the integrity of the detector as electrical discharges may cause irreversible damages to GEM foils or readout electronics.

Spark discharges are associated with the development of a streamer after exceeding a critical charge in a single GEM hole~\cite{Bressan, Mathis_GEMpaper} and were extensively studied in single- and multi-GEM structures ~\cite{Bachmann}. Several recommendations for a safe operation of GEM-based detectors have been worked out. These include the introduction of protection resistors, the reduction of the active area of single GEM segments, and the application of optimized potentials to the subsequent electrodes of a multi-GEM structure \cite{Bachmann}. 

However, it was observed that a primary spark discharge in a GEM hole may trigger a secondary discharge in the gap below the GEM~\cite{Bachmann, Peskov}. The latter occur already at electric field values lower than the amplification field in the given gas mixture with a delay $\mathcal{O}$(\textmu s) \cite{Deisting:2019qda}. As the secondary discharges appear to be more violent than the primary ones, a mitigation of the former is essential for the stable operation of GEM-based detectors. It was recently shown that the secondary discharge probability drops when the potential of the GEM bottom electrode is defined through a non-zero resistance
$\mathcal{O}$(10-100\,k$\Omega$)~\cite{Deisting:2019qda, Utrobicic:2019tss}. While the exact mechanism of secondary discharge creation is still a subject of debate (including the most recent findings by \cite{Deisting:2019qda, Utrobicic:2019tss}, where thermionic emission from the cathode is proposed as a possible explanation of their development) it is important to study all possible ways of mitigation of these violent events.  

In this manuscript we intend to compile a set of further recommendations to mitigate the occurrence of secondary discharges by optimizing the $RC$ characteristics of the detector and the high voltage (HV) power supply scheme. This work systematically investigates the secondary discharge probability in single GEM detectors as a function of the field below the GEM varying $RC$ components of the setup. The paper is structured as follows. The experimental setup is introduced in Sec.~\ref{sec:setup}. In Sec.~\ref{sec:1GEM} the single GEM measurements are presented and their results discussed. Section~\ref{sec:conclusion} summarizes the findings and concludes the paper.

%% file: Setup.tex
\section{Experimental Setup}
\label{sec:setup}
\begin{figure}[htbp]
\centering
\includegraphics[width=0.9\columnwidth]{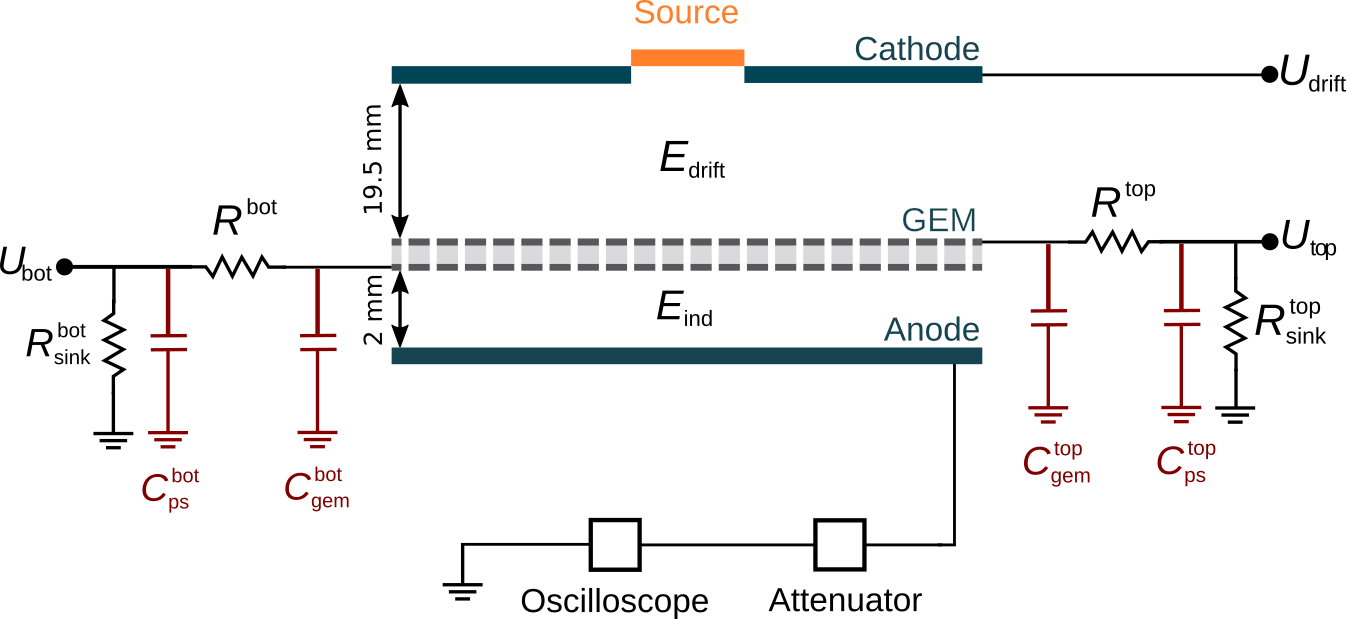}
\caption{Schematic of the detector and its powering scheme. A single GEM is mounted between a drift electrode and a readout anode. The anode is read out by an oscilloscope via an attenuator. The relevant parasitic capacitances introduced by cables are marked in red. See text for more details.}
\label{fig:setup}
\end{figure}

A dedicated detector setup was built to study the influence of the $RC$ elements on the secondary discharge probability. Figure~\ref{fig:setup} shows a view of the setup and its powering scheme.

The detector vessel contains a $10\times10$\,cm$^2$ GEM foil mounted on a 2\,mm thick FR4 frame with a readout anode below and a drift cathode above the GEM. Both cathode and anode are made of a 1.5 mm thick PCB coated with copper on one side.
The standard GEM foil, produced by TECHTRA sp.\,z\,o.o in Poland~\cite{Techtra} with the double-mask technology, consists of a 50\,\textmu m thick polyimide (Apical) foil covered on both sides with a 5\,\textmu m copper layer, perforated with holes with 50\,\textmu m inner and 70\,\textmu m outer hole diameter at a pitch of 140\,\textmu m.
The distance between the cathode and the GEM (drift gap) is 19.5\,mm. The distance between the GEM and the anode (induction gap) is set to 2\,mm throughout all measurements.

Potentials are applied using a HV power supply with independent channels.
The drift cathode potential ($U_{\mathrm{drift}}$) is applied directly to the cathode, whereas GEM potentials ($U_{\mathrm{top}}$ and $U_{\mathrm{bot}}$) are defined through the protection resistors ($R^{\mathrm {top}}$ and $R^{\mathrm {bot}}$) on the GEM top and bottom electrodes, respectively. The GEM potentials are varied within the course of these studies to investigate their effect on the secondary discharge probability.

To ensure safe and fast discharge of the GEM in case of a power supply trip, resistors to ground $R^{\mathrm {top}}_{\mathrm{sink}}$ and $R^{\mathrm {bot}}_{\mathrm{sink}}$ -- 5\,M$\Omega$ at the top side, 10\,M$\Omega$ at the bottom side -- are installed. They are kept constant for all the systematic studies shown in this manuscript. 

In this work, the influence of the cable length on the secondary discharge probability is studied. All cables connecting HV scheme components with the detector introduce a parasitic capacitance to the system, hence they are marked in Fig.~\ref{fig:setup} as four capacitors to ground. $C^{\mathrm {top}}_{\mathrm{ps}}$ and $C^{\mathrm {bot}}_{\mathrm{ps}}$ correspond to the cables connecting the power supply with the top and bottom protection resistors, respectively. The parasitic capacitances $C^{\mathrm {top}}_{\mathrm{gem}}$ and $C^{\mathrm {bot}}_{\mathrm{gem}}$ are introduced by the cables between the protection resistors and the GEM top and bottom electrodes. The cables used in the setup are typical coaxial and shielded HV cables with parasitic capacitance of about 100\,pF per meter.

The detector volume is flushed with either Ar-CO$_{2}$ (90-10) or Ne-CO$_{2}$-N$_{2}$ (90-10-5)\footnote{ALICE TPC gas mixture} at atmospheric pressure. The oxygen and water concentrations are constantly monitored and it is ensured that the oxygen level is below 20 ppm and the water contamination is in the ppm region.

The detector is operated with a drift field ($E_{\mathrm{drift}}$), defined by $U_{\mathrm{drift}}$ and $U_{\mathrm{top}}$, of 400\,V/cm. To induce discharges a $^{239}$Pu+$^{241}$Am+$^{244}$Cm $\alpha$-emitter \cite{alpha} is mounted on top of a centered hole in the cathode.
The voltage across the GEM ($\Delta U_{\mathrm{GEM}} = U_{\mathrm{top}}-U_{\mathrm{bot}}$) is chosen to achieve gains high enough to measure primary discharge rates of $\SIrange{0.1}{1.0}{}$\,Hz. This corresponds to typical gains of $\sim$315 for Ar-CO$_{2}$ (90-10) and $\sim$1700 for Ne-CO$_{2}$-N$_{2}$ (90-10-5).
The induction field ($E_{\mathrm{ind}}$) below the GEM is varied throughout all measurements to measure its influence on the creation of secondary discharges.

\begin{figure}[!tbp]
 \centering
 \includegraphics[width=0.9\columnwidth]{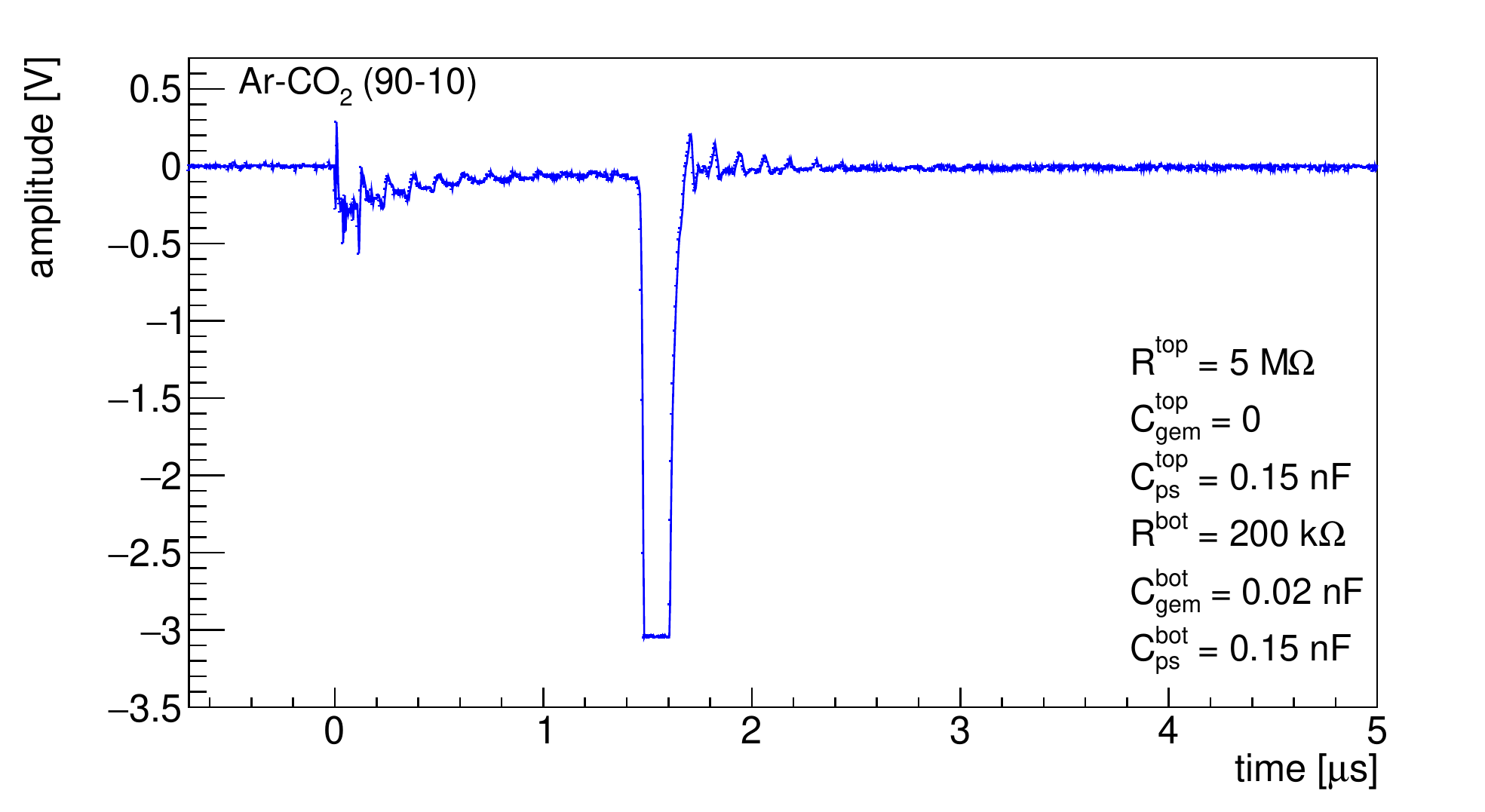}
        \caption{Signal of a primary discharge followed by a secondary discharge measured at $E_{\mathrm{ind}} = 5500$\,kV/cm with a 34\,dB attenuator. Signal of the secondary discharge over-saturates the oscilloscope channel. Here and in all following figures, it is assumed that 1\,m of cable corresponds to 0.1\,nF capacitance.}
    \label{fig:prim_highAmp}
\end{figure}

The anode is read out by an oscilloscope via a $\SIrange{31}{34}{}$\,dB attenuator, to avoid saturation of the channel. A typical waveform of a primary discharge, followed by a secondary, is shown in Fig.~\ref{fig:prim_highAmp}. In this case, signal of the secondary discharge appears $\sim$1.4\,\textmu s after the primary and has a larger amplitude making it easy to distinguish between the two. Experience shows, however, that the secondary discharge amplitude and the time between primary and secondary discharge strongly depends on the applied $E_{\mathrm{ind}}$ and the specific HV scheme components. It may happen that the amplitude of the secondary is lower than the amplitude of the primary discharge signal. Thus, readout based on discriminators and scalers is not feasible and an event-by-event waveform analysis needs to be considered. In the studies we use the Yokagawa DLM 2054 oscilloscope, which allows to store waveforms at a rate of $\sim$10\,kHz with a 100\% efficiency. The waveforms are analyzed offline and the number of primary and secondary discharges for a given measurement is extracted. The probability ${P}_{\mathrm {sec}}$ for secondary discharge to occur is defined as the ratio of the number of secondary to primary discharges. 

%% file: SingleGEM.tex
\section{Secondary discharge mitigation studies}
\label{sec:1GEM}

In this section the influence of the induction gap uniformity and $RC$ components of the HV powering scheme on the secondary discharge probability is discussed.

\subsection{Influence of mechanical support}
\begin{figure}[!htbp]
\centering
\includegraphics[width=0.9\columnwidth]{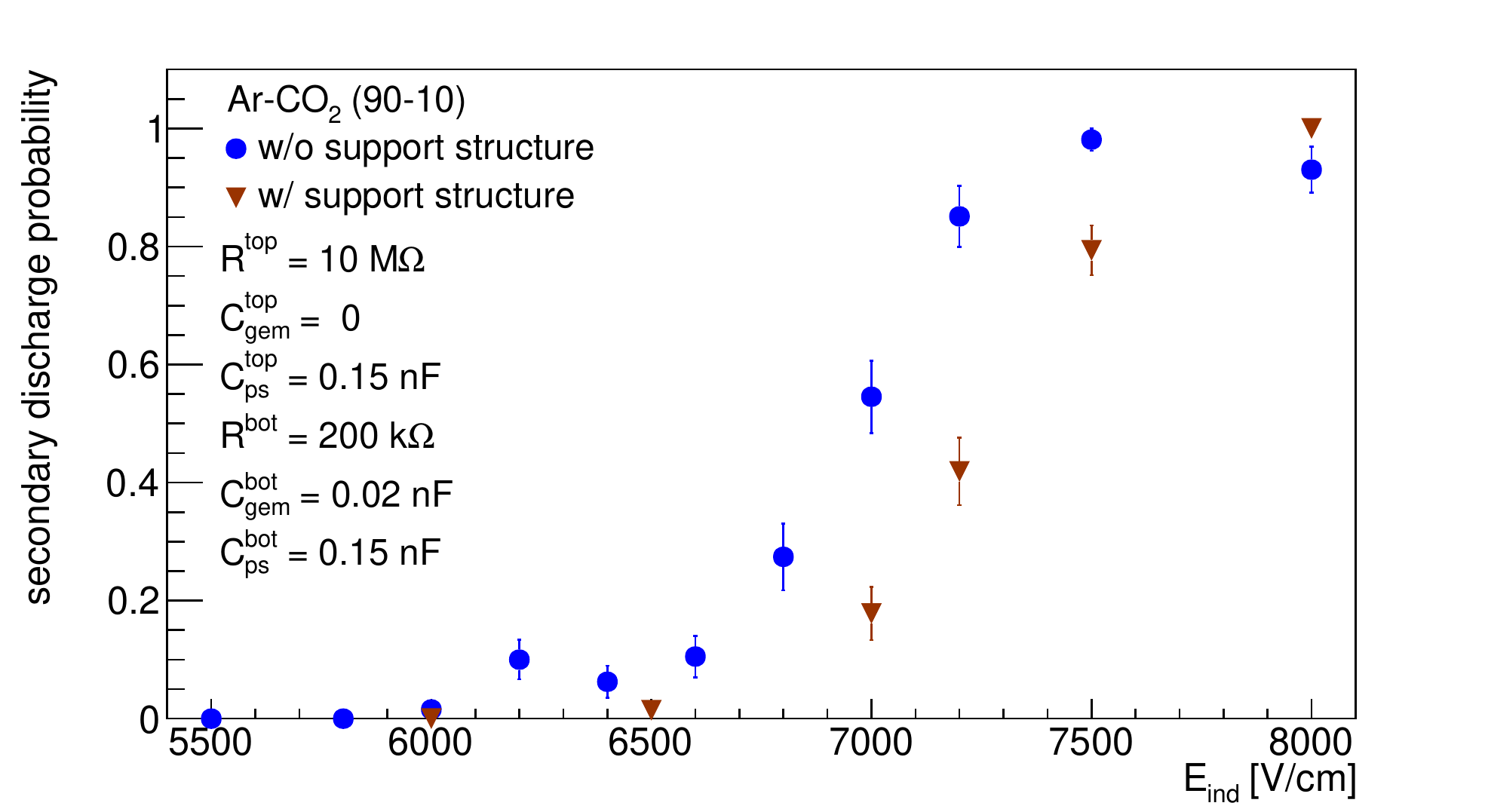}
\caption{Secondary discharge probability as a function of the induction field measured with and without the support structure in the induction gap.}
\label{fig:cross}
\end{figure}

Although the mechanism of secondary discharges cannot be explained simply by the increasing the voltage (and thus electric field) across the gap below the discharging GEM \cite{Deisting:2019qda, Utrobicic:2019tss}, it is necessary to keep this field constant after a primary spark, as the secondary discharge probability strongly depends on its value. One possible effect to be considered is the bending of a GEM foil towards the readout anode due to gravity or electrostatic attraction. Figure~\ref{fig:cross} shows a comparison of secondary discharge probability measured with a GEM mounted with and without additional support underneath to prevent the foil from sagging. The support structure is a cross consisting of two equally long arms with 1.4\,mm width and 2\,mm thickness made out of G11 composite material covering $\sim$3\% of the active GEM area.

The measurements are performed in Ar-CO$_2$ (90-10). Values of the $RC$ components, described in detail in Sec.~\ref{sec:setup}, are given in the figure legend. Parasitic capacitance values correspond to the cable lengths assuming 100\,pF per meter. It is shown that with the support structure the onset field of secondary discharges shifts towards higher values, however, the effect is not significant enough to explain the development of secondary discharges, which in both cases appear at the fields far below the amplification field value in the used gas mixture ($\sim$10\,kV/cm). Nevertheless, one should consider applying sufficient stretching force while mounting GEMs, installing a support structure below large-area foils, or increasing the gap widths when designing and constructing a detector. 

\subsection{Influence of \texorpdfstring{$C^{\mathrm{top}}_{\mathrm{gem}}$}{ctopgem}capacitance}
\label{sec:ctopgem}
\begin{figure}[!htbp]
\centering
\includegraphics[width=0.9\columnwidth]{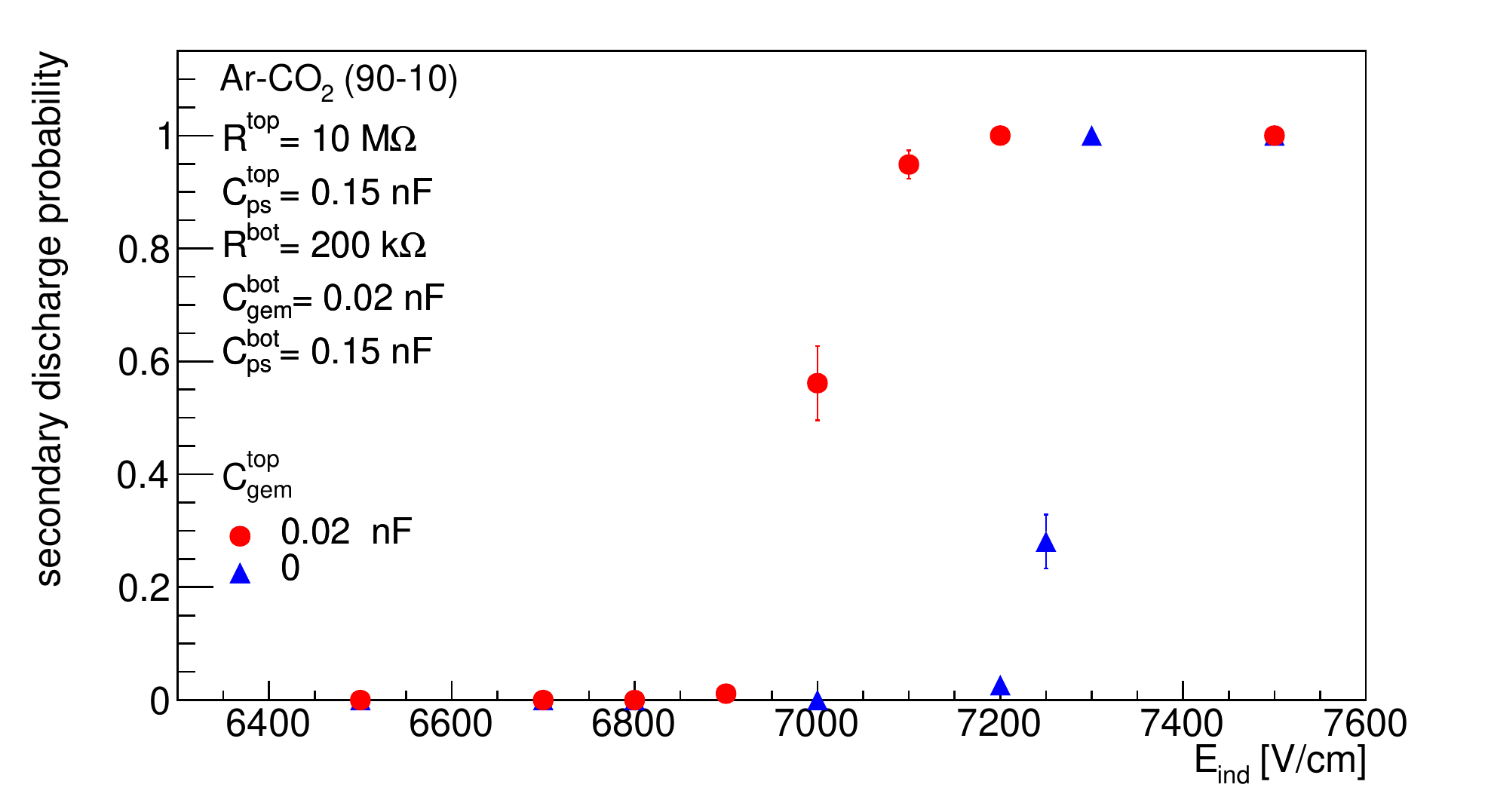}
\caption{Secondary discharge probability as a function of the induction field for different cable lengths between the $R^{\mathrm {top}}$ resistor and the GEM top electrode.}
\label{fig:C_top}
\end{figure}

In the following, the connection between the $R^{\mathrm {top}}$ resistor and the GEM top electrode is studied. Two configurations are used: in the first one $R^{\mathrm {top}}$ is soldered directly to the electrode, in the second one a parasitic capacitance in the form of a 20\,cm long cable between $R^{\mathrm {top}}$ and the GEM top electrode is introduced. The results are presented in Fig.~\ref{fig:C_top}. The additional capacitance shifts the onset of secondary discharges towards lower field values  by $\SIrange{300}{400}{}$\,V/cm. This is due to the extra energy stored in the HV cable, which can be released during a primary discharge. Effectively, the $U_\mathrm{bot}$ potential (and thus the induction field) increases after a primary discharge making an appearance of a secondary discharge more likely. It has already been shown that the secondary discharge probability increases with the energy stored in the foil  hence energy of the primary discharge  \cite{Bachmann}. Thus, all the cables supplying potentials to the GEM electrodes (top-side in this particular study) should be decoupled by a high-ohmic protection resistor (see Sec.~\ref{sec:rtop} for more details) mounted directly on the foil, or their lengths shall be reduced to the minimum.

\subsection{Influence of \texorpdfstring{$R^{\mathrm{top}}$}{rtop} resistance}
\label{sec:rtop}
Usually, a protection resistor $R^{\mathrm{top}}$ is mounted on a GEM in order to quench primary discharges, which may appear during detector operation, and protect the foil from damages. In segmented GEM foils the $R^{\mathrm{top}}$ also reduces the current flowing through the GEM in case of a short circuit in one of the segments. The typical values of the protection resistance vary between 1 and 10\,M$\Omega$, depending on the maximum voltage (and thus gain) drop allowed by the experimental requirements or number of GEM segments with a permanent short which can be handled by a HV supply system.

\begin{figure}[!htbp]
\centering
\includegraphics[width=0.9\columnwidth]{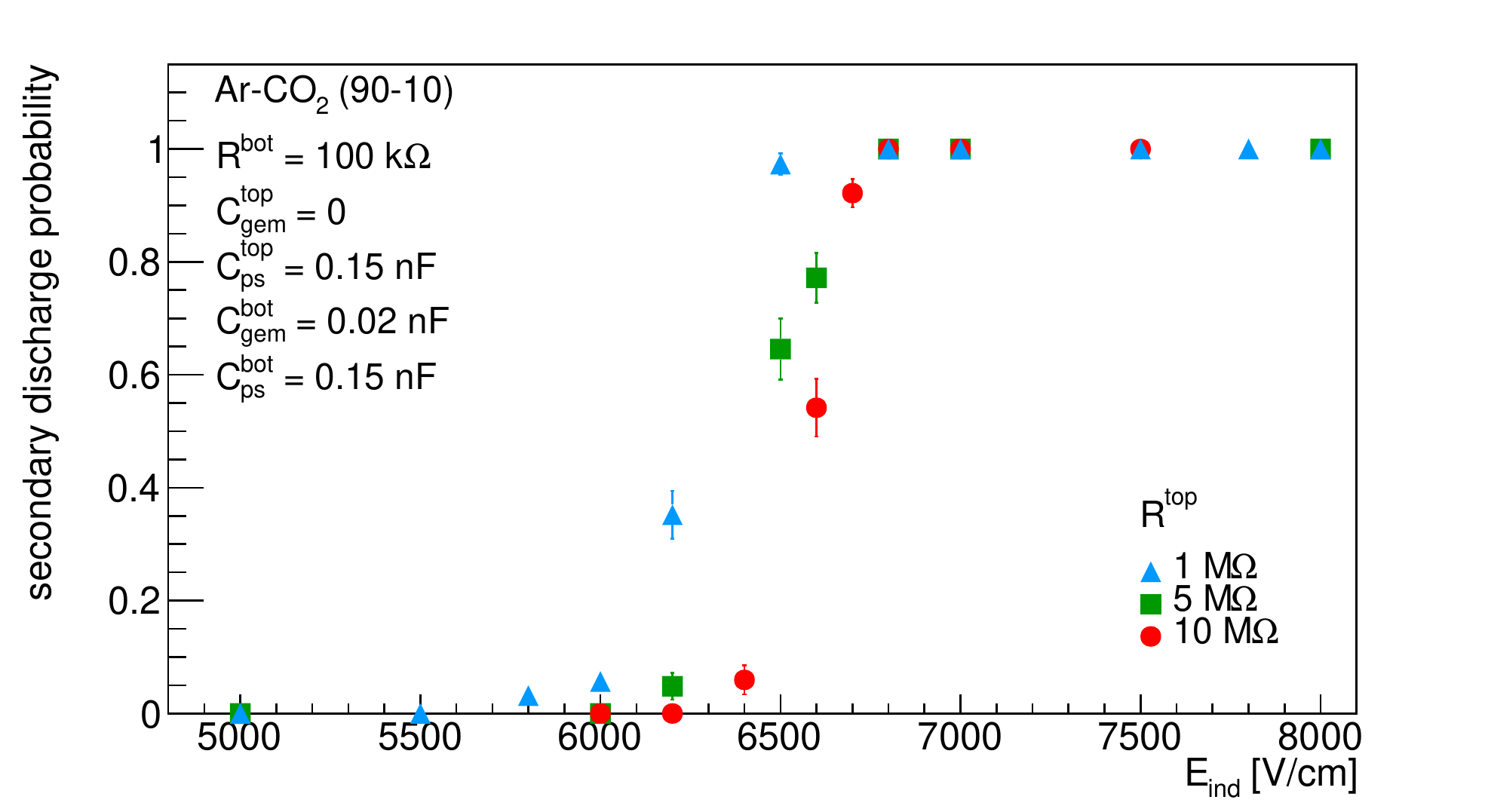}
\caption{Secondary discharge probability as a function of the induction field for different  $R^{\mathrm {top}}$ values.}
\label{fig:Rtop}
\end{figure}

However, as shown in Fig.~\ref{fig:Rtop}, where the secondary discharge probability is measured for different $R^{\mathrm {top}}$ values, the $P_{\mathrm{sec}}$ does not significantly depend on the choice of $R^{\mathrm {top}}$, although higher $R^{\mathrm {top}}$ slightly shift the onset towards higher field strength. This is related to the fact that extra charge can be fed into a primary discharge from the power supply, increasing its energy, when running with a lower value of $R^{\mathrm {top}}$. In addition, a higher value of $R^{\mathrm {top}}$ with respect to $R^{\mathrm {bot}}$ assures a smaller increase of the field below the discharging GEM and thus, reduces the secondary discharge probability. However, the recommendation of setting the $R^{\mathrm {top}}$ resistance as high as possible must be balanced with the maximum gain drop allowed by the experimental requirements. 

\subsection{Influence of \texorpdfstring{$C^{\mathrm{top}}_{\mathrm{ps}}$}{ctopps} capacitance}
\label{sec:ctopps}
\begin{figure}[!htbp]
\centering
\includegraphics[width=0.9\columnwidth]{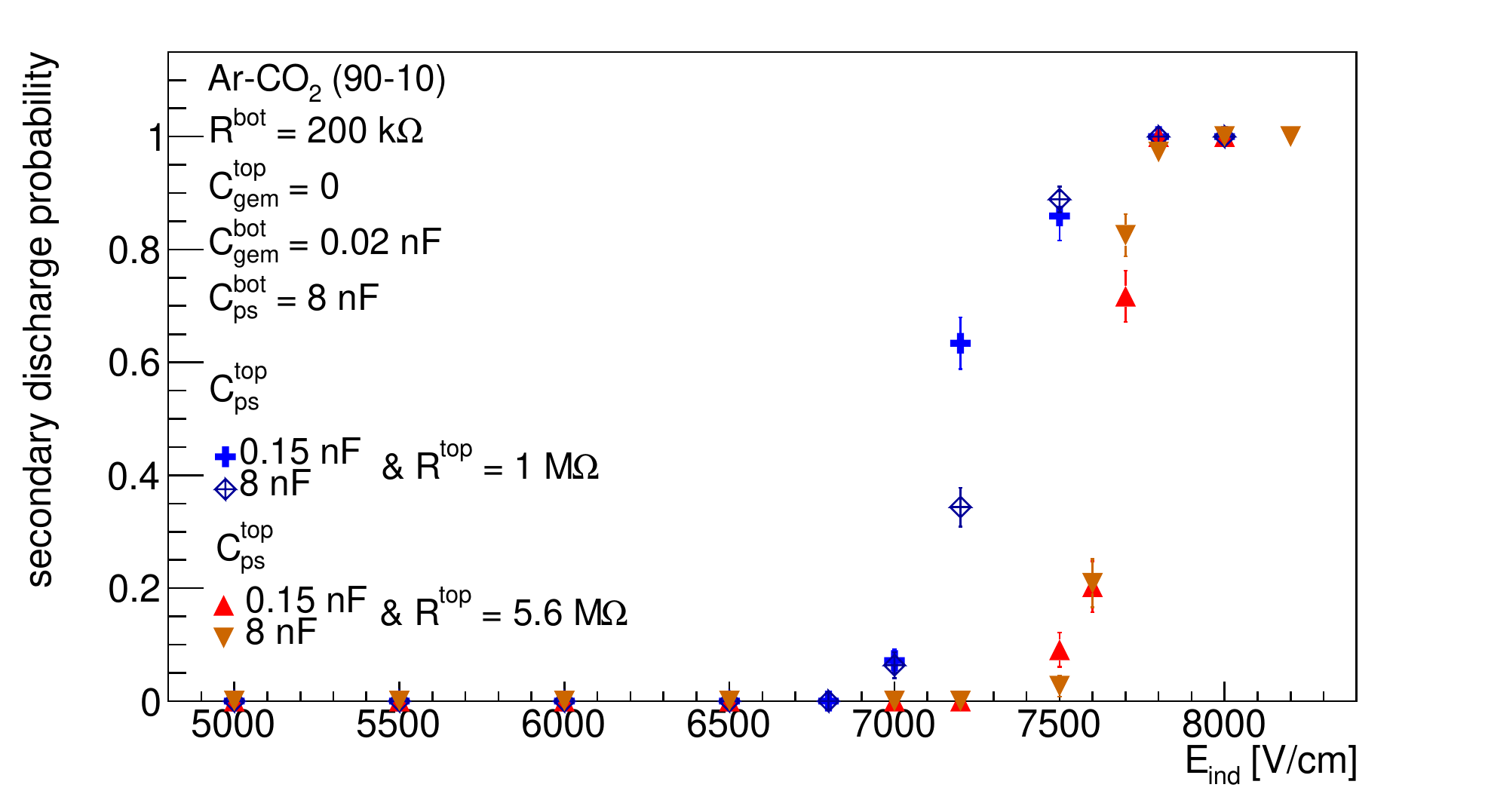}
\caption{Decoupling effect of $R^{\mathrm {top}}$ measured for different cable lengths between the HV power supply channel and the top protection resistor.}
\label{fig:decoupling_Rtop}
\end{figure}

The influence of the $C_{\mathrm{ps}}^{\mathrm{top}}$ capacitance is studied by changing the length of the cable providing high voltage from the power supply. However, the capacity of the cable (and thus stored energy) is decoupled from the GEM detector using the $R^{\mathrm {top}}$ resistor. Figure~\ref{fig:decoupling_Rtop} shows the results for two different cable lengths (1.5\,m and 80\,m corresponding to $C_{\mathrm{ps}}^{\mathrm{top}}$ of 0.15\,nF and 8\,nF) decoupled by 1\,M$\Omega$ and 5.6\,M$\Omega$ $R^{\mathrm {top}}$ resistors. For both resistances, there is no significant difference in the onset fields for secondary discharges. This means that the typical $R^{\mathrm {top}}$ values of $\mathcal{O}$(1\,M$\Omega$) are sufficient to minimize the influence of the extra capacitance (e.g. HV supply cable). The difference in the secondary discharge probability measured for two values of $R^{\mathrm {top}}$ confirms the results presented in Sec.~\ref{sec:rtop}. Here, however, the measurements were performed for a higher value of $R^{\mathrm{bot}}$, which shifts the onset of secondary discharges towards higher fields. The influence of $R^{\mathrm {bot}}$ on the secondary discharge probability is studied in detail in the following section.

\subsection{Influence of \texorpdfstring{$R^{\mathrm{bot}}$}{rbot} resistance}
\label{sec:rbot}
The employment of the bottom resistance $R^{\mathrm{bot}}>0$ is crucial for the mitigation of secondary discharges \cite{Deisting:2019qda}. It was shown that a primary discharge in a GEM is followed by a current development in the gap below the discharging foil \cite{Deisting:2019qda, Utrobicic:2019tss}. This induces currents in the surrounding electrodes, i.e. bottom GEM and readout anode in case of a single-GEM setup. If a resistor is connected in series to one or both of these electrodes, the current causes a potential drop which leads to a reduction of the electric field in the gap below the discharging GEM (e.g. induction gap in a single-GEM setup). Due to the strong dependence of the secondary discharge probability on the field strength, a reduction of the latter quenches the development of secondary discharges. This effect is much stronger than the slight increase of the field caused by a discharging capacitor (GEM foil) with non-zero resistances connected in series to both, top and bottom, electrodes \cite{Deisting:2019qda}. 

\begin{figure}[!htbp]
\centering
\includegraphics[width=0.9\columnwidth]{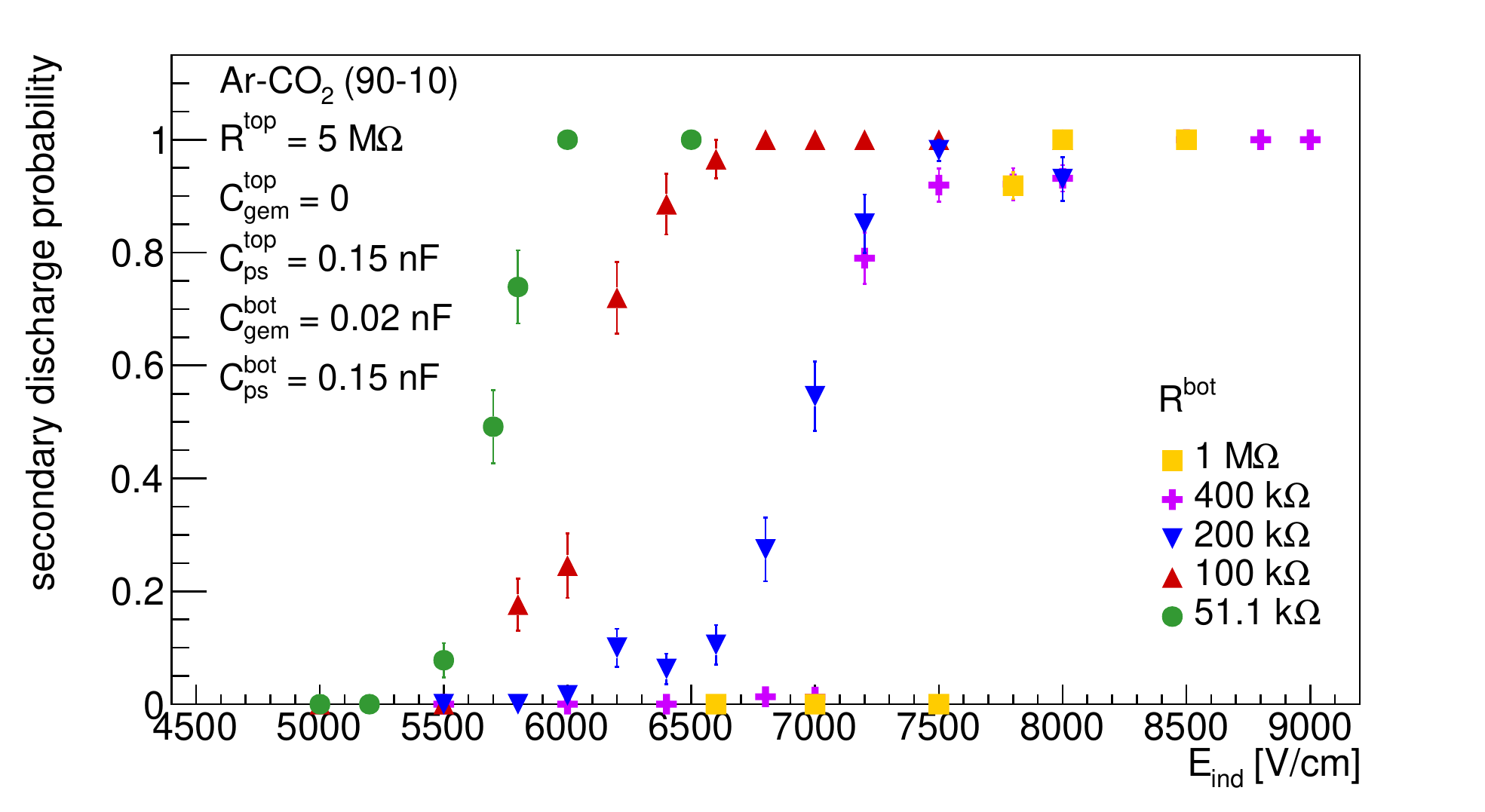}
\caption{Secondary discharge probability as a function of $E_{\mathrm{ind}}$ measured for different values of $R^{\mathrm {bot}}$ in Ar-CO$_2$ (90-10).}
\label{fig:Rbot_Ar}
\end{figure}

\begin{figure}[!htbp]
\centering
\includegraphics[width=0.9\columnwidth]{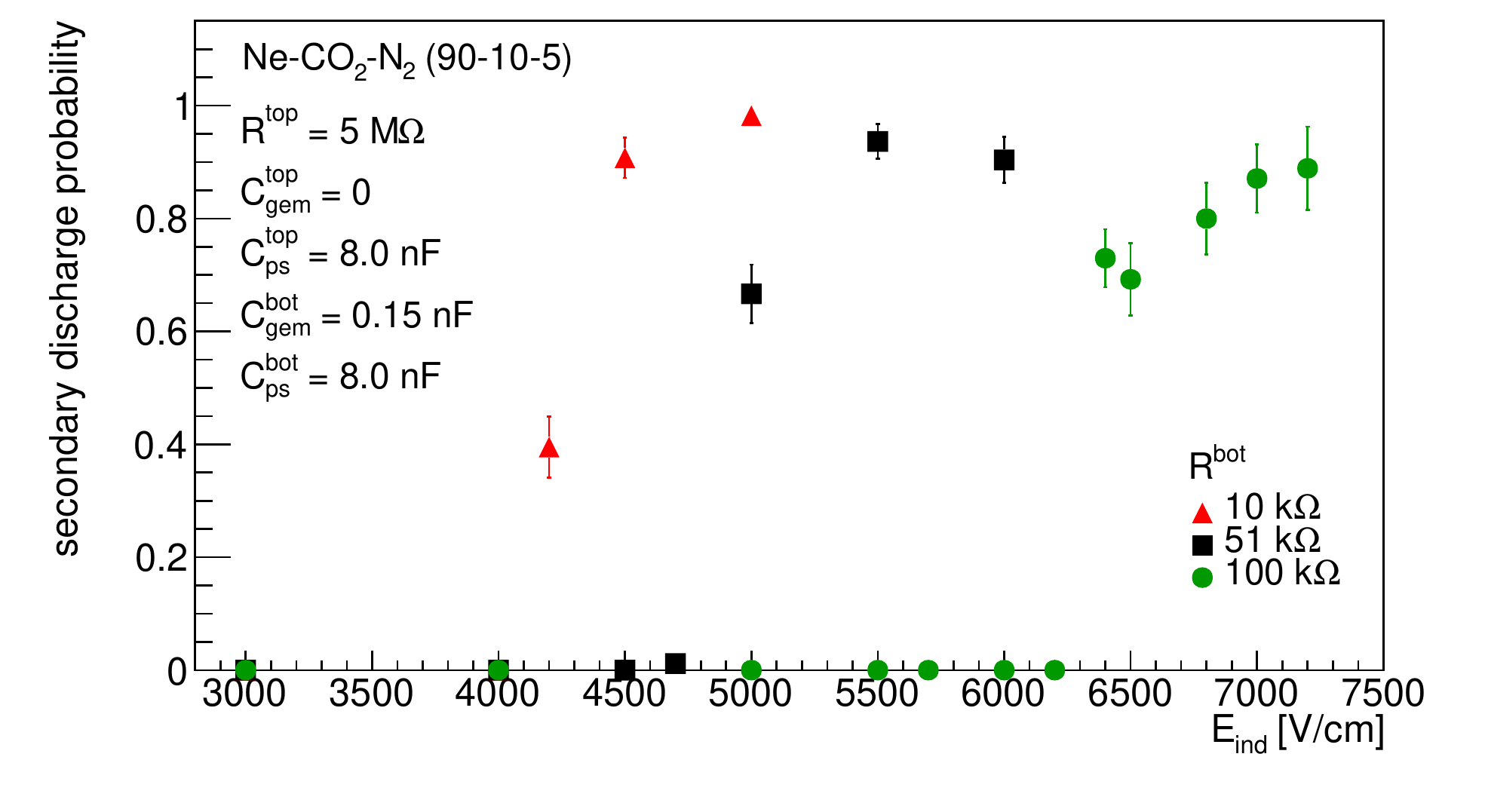}
\caption{Secondary discharge probability as a function of $E_{\mathrm{ind}}$ measured for different values of $R^{\mathrm {bot}}$ in Ne-CO$_2$-N$_2$ (90-10-5).}
\label{fig:Rbot_Ne}
\end{figure}

Figures~\ref{fig:Rbot_Ar} and~\ref{fig:Rbot_Ne} show the secondary discharge probability measured as a function of $E_{\mathrm{ind}}$ for various values of $R^{\mathrm {bot}}$ in  Ar-CO$_2$ (90-10) and Ne-CO$_2$-N$_2$ (90-10-5) gas mixtures. In both gases a clear trend can be observed: the larger $R^{\mathrm {bot}}$ the higher is the induction field at which secondary discharges start to occur.
In neon, however, the secondary discharges start occurring at lower fields than in argon due to the higher value of the Townsend coefficient in the former. It should be noted that also in this case the onset field for secondary discharges is lower than the amplification field ($\sim$5\,kV/cm in Ne-CO$_2$-N$_2$ (90-10-5), $\sim$7\,kV/cm in Ar-CO$_2$ (90-10) \cite{Deisting:2019qda}). This clearly shows that the mechanism of secondary discharge creation is more complicated than a simple charge amplification in the gas.

To summarize, in order to minimize the secondary discharge occurrence it is recommended to consider higher values of $R^{\mathrm{bot}}$ resistance while optimizing the HV scheme of the GEM system. This, of course, needs to be balanced with the contradicting effect of the potential (hence gain) drop across this resistor during operation.

\subsection{Influence of \texorpdfstring{$C^{\mathrm{bot}}_{\mathrm{gem}}$}{cbotgem} capacitance}
\label{sec:cbotgem}

\begin{figure}[!htbp]
\centering
\includegraphics[width=0.9\columnwidth]{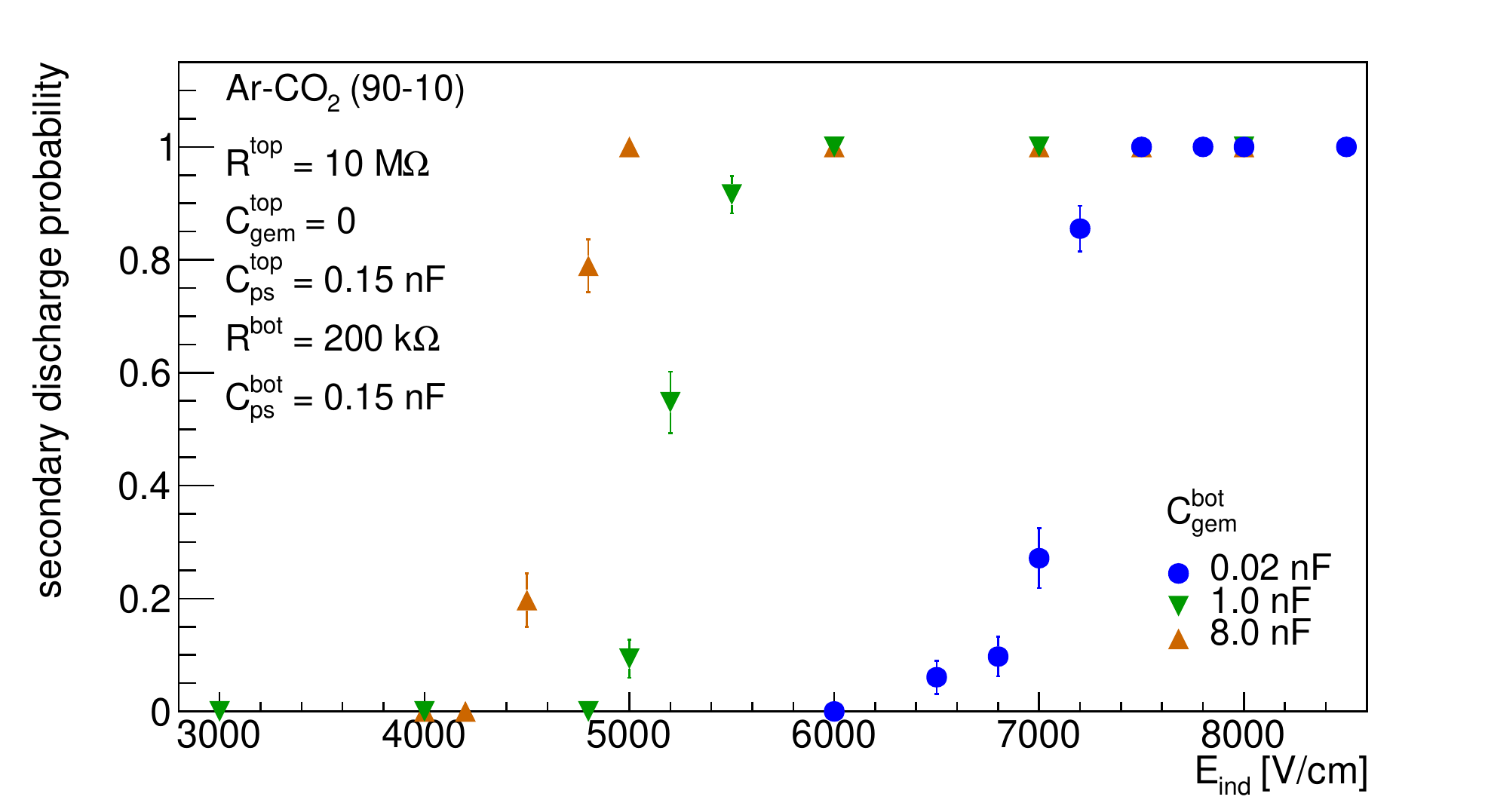}
\caption{Secondary discharge probability as a function of $E_{\mathrm{ind}}$ for different cable lengths between $R^{\mathrm {bot}}$ and the GEM bottom electrode.}
\label{fig:plot_C_between_Ar}
\end{figure}

Studying the influence of the cable length between the GEM bottom electrode and the resistor $R^{\mathrm {bot}}$ on the secondary discharge probability, it becomes apparent that with higher parasitic capacitance the secondary discharges occur at lower induction fields (see Fig.~\ref{fig:plot_C_between_Ar}). 
The cable introduces an extra capacitance to the ground in parallel to the capacitance of the induction gap. Hence, more energy is stored in the system, which is then released in the process of the secondary discharge. The effective $E_{\mathrm{ind}}$ after a primary discharge is higher and thus a secondary discharge is more likely to occur.
Therefore, to increase the stability of a GEM detector it can be recommended to reduce the capacitance between the GEM bottom electrode and $R^{\mathrm {bot}}$. This can be done by keeping the cables short or even soldering the protection resistor directly to the GEM bottom electrode. 

However, note, that the capacitance of the induction gap is defined by the size of the GEM foil. In case of large-area detectors the effect shall be the same as for the large parasitic capacitance $C^{\mathrm{bot}}_{\mathrm{gem}}$. A special care should be given to the design of large-area systems, including segmentation of the bottom electrode. Also, increasing the distance between GEMs (or in case of the induction gap, between the GEM bottom electrode and the readout anode) shall improve stability significantly.

\subsection{Influence of \texorpdfstring{$C^{\mathrm{bot}}_{\mathrm{ps}}$}{cbotps} capacitance}
\label{sec:cbotps}
\begin{figure}[!htbp]
\centering
\includegraphics[width=0.9\columnwidth]{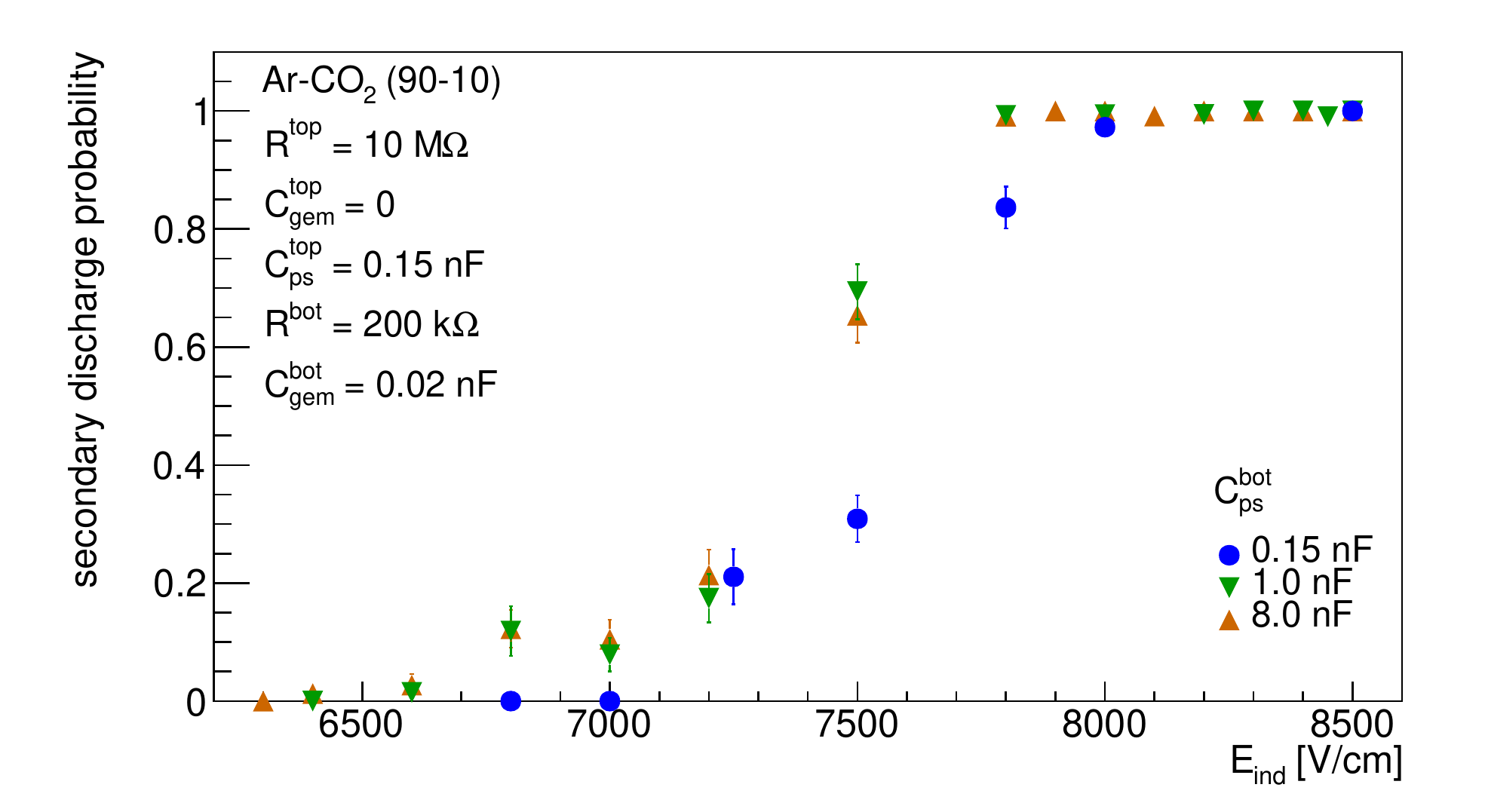}
\caption{Secondary discharge probability as a function of $E_{\mathrm{ind}}$ for different cable lengths between the bottom protection resistor and the HV power supply channel, measured in Ar-CO$_2$ (90-10).}
\label{fig:C_after_Ar}
\end{figure}

In the following, the influence of the cable length between the power supply and the decoupling resistor $R^{\mathrm {bot}}$ on the secondary discharge probability is investigated.
Figure~\ref{fig:C_after_Ar} presents the results obtained in Ar-CO$_2$ (90-10) with $R^{\mathrm{bot}} = 200$\,k$\Omega$. Similar to the top-electrode considerations (see Sec.~\ref{sec:ctopps}), with an $R^{\mathrm {bot}}$ resistance in the order of 100\,k$\Omega$ the parasitic capacitance introduced by a cable does not influence the secondary discharge probability. In the discussed example, all three capacitances (corresponding to 1.5\,m, 10\,m, and 80\,m cables) are well decoupled from the GEM detector.

\begin{figure}[!htbp]
 \centering
 \includegraphics[width=0.9\columnwidth]{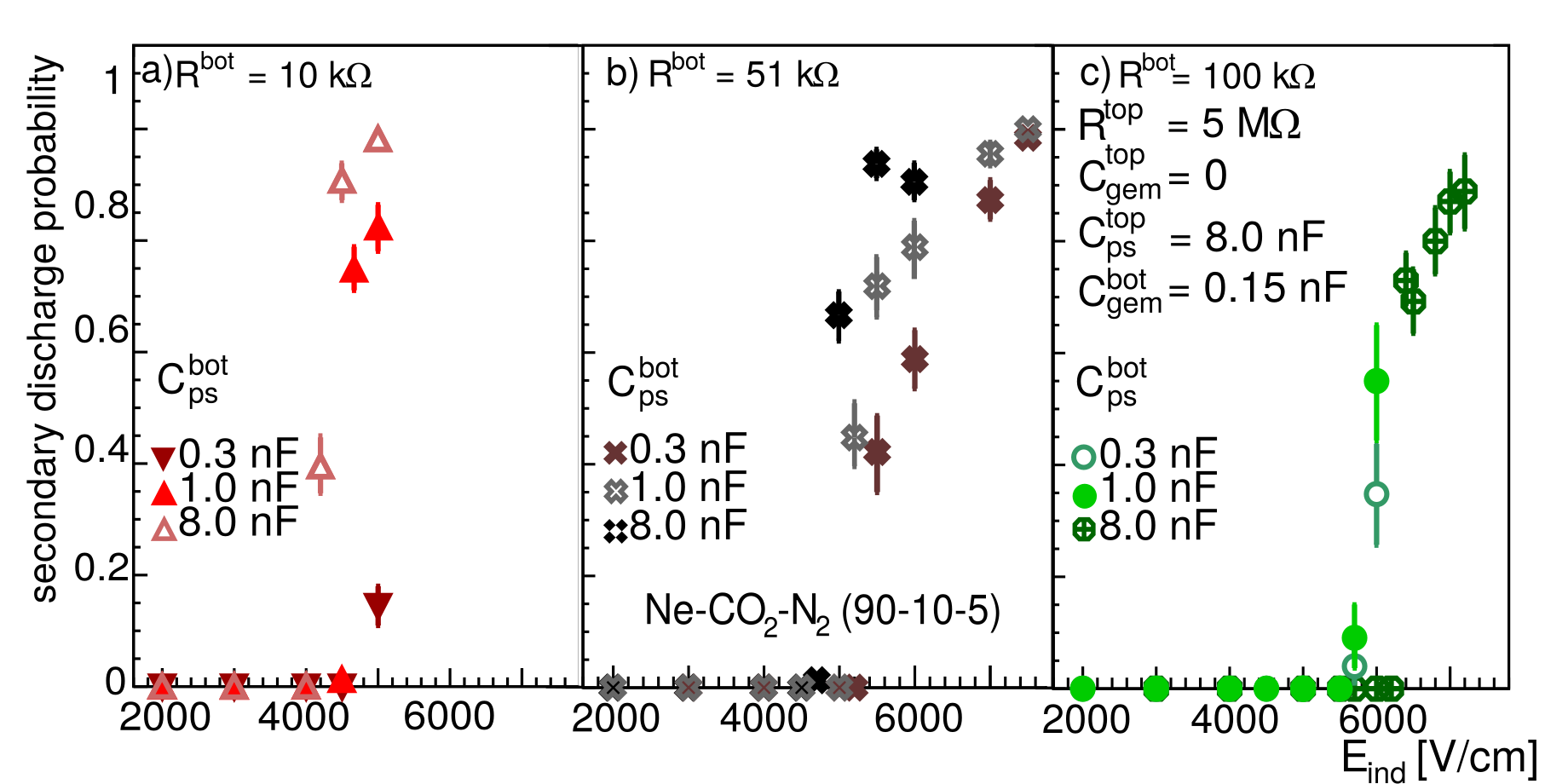}

\caption{Secondary discharge probability as a function of $E_{\mathrm{ind}}$ for different cable lengths between the bottom protection resistor and the HV power supply channel, measured in Ne-CO$_2$-N$_2$ (90-10-5). Plots in different panels correspond to different values of the protection resistance $R^{\mathrm{bot}}$.}
 \label{fig:decoupling_Ne}
 \end{figure}

Measurements in Ne-CO$_2$-N$_2$ (90-10-5) using different cable lengths (3\,m, 10\,m, and 80\,m) performed with an $R^{\mathrm {bot}}$ resistance of 10\,k$\Omega$, 51\,k$\Omega$, and 100\,k$\Omega$ are presented in Fig.~\ref{fig:decoupling_Ne}. Following the discussion in Sec.~\ref{sec:rbot}, the onset fields for secondary discharges shift towards higher values with increasing $R^{\mathrm {bot}}$ resistance. In addition, measurements with low-value resistors (panel \textit{a} and \textit{b}) reveal a dependency on the cable length. Secondary discharges occur at about 1000\,V/cm lower fields when a long cable is connected. Similar to $C^{\mathrm{bot}}_{\mathrm{gem}}$ considerations (see Sec.~\ref{sec:cbotgem}), the cable functions as an extra energy reservoir. This dependency disappears when $R^{\mathrm {bot}}=100$\,k$\Omega$ is connected in series between the power supply and the GEM bottom electrode (panel \textit{c}). 

%% file: Conclusion.tex
\section{Conclusion}
\label{sec:conclusion}

The paper summarizes our findings about a safe HV scheme for a GEM-based detector studied with a single-GEM, $10\times10$\,cm$^2$ setup. Parasitic capacitances introduced by the coaxial cables and protection resistors for both, GEM top and bottom, electrodes were studied.

From all the studies presented here and in previous works a clear secondary discharge mitigation strategy emerges, which involves the reduction of the transfer and induction fields in the system. This is important especially for gas mixtures where amplification starts at the field values close to the secondary discharge onset, e.g. in neon-based mixtures. The field considerations can be summarized in the following list of recommendations: 
\begin{itemize}
    \item Transfer and induction fields should be reduced below the secondary discharge onset level, if compatible with the operation requirements.
    \item Transfer and induction gaps in the detector should be kept uniform to avoid electric field enhancement. Using support structures is recommended, when possible.
\end{itemize}

In addition, the optimization of the HV scheme should address the following points in order to mitigate a risk of secondary discharge occurrence.  
\begin{itemize}
    \item The protection resistors, connected in series between a GEM electrode and a HV source, should be installed as close to the GEM electrode as possible. If use of a cable cannot be avoided, its length (and thus capacitance) should be minimized. This concerns both, top and bottom, sides of a foil.
    \item For the same reason it is recommended to reduce the capacitance of the gaps by increasing the gap size, limiting single detector area or segmenting the bottom side of a GEM foil.
    \item Typical values of $\SIrange{1}{10}{}$\,M$\Omega$ for the top-side protection resistors are well in line with the mitigation strategy.    
    \item The non-zero bottom-side protection resistor is crucial for the secondary discharge mitigation. Values in the order of 100\,k$\Omega$ are found to efficiently reduce the secondary discharge probability.
    \item Length of the cables providing electric potential from the HV source do not influence the stability as long as properly decoupled by the protection resistors (see values above).
\end{itemize}
    
However, it should be noted that for each detector design, a careful optimization should be carried out to take into account project-specific constraints. The results presented in this work were obtained for a small-area, single-GEM setup. In multi-GEM detectors an interplay between GEM voltages and transfer gaps may introduce extra effects, which may put additional constraints to the lists of recommendations above (see recent findings by the CMS Muon Group \cite{MerlinMPGD}).

Further studies on the optimization of the HV scheme are planned by the authors, including systematic measurements with multi-GEM structures and resistive electrodes (GEM, readout boards). The latter may be an efficient way to mitigate secondary discharges providing internal quenching capabilities of such structures.

%% file: main.bbl
\begin{thebibliography}{99}

\bibitem{Sauli}F.~Sauli,
  \emph{GEM: A new concept for electron amplification in gas detectors},
  Nucl.\ Instrum.\ Meth.\ A 386 (1997) 531
  doi:10.1016/S0168-9002(96)01172-2
  
\bibitem{compass_Altunbas}
C.~Altunbas, M.~Cape\'{a}ns, K.~Dehmelt, J.~Ehlers, J.~Friedrich, I.~Konorov, A.~Gandi, S.~Kappler, B.~Ketzer, R.~Oliveira,S.~Paul, A.~Placci, L.~Ropelewski, F.~Sauli, F.~Simon and M.~van Stenis, \emph{Construction, test and commissioning of the triple-gem tracking detector for compass},Nucl.\ Instrum.\ Meth.\ A 490 (2002) 177, 10.1016/S0168-9002(02)00910-5. 

\bibitem{compass_Ketzer}
B.~Ketzer, Q.~Weitzel, S.~Paul, F.~Sauli and L.~Ropelewski \emph{Performance of triple GEM tracking detectors in the COMPASS experiment}, Nucl.\ Instrum.\ Meth.\ A 648 (2004) 314,
10.1016/j.nima.2004.07.146. 

\bibitem{LHCb_Bencivenni}
G.~Bencivenni, G.~Felici, F.~Murtas, P.~Valente, W.~Bonivento, A.~Cardini, A.~Lai, D.~Pinci, B.~Saitta and C.~Bosio, \emph{A triple GEM detector with pad readout for high rate charged particle triggering}, Nucl.\ Instrum.\ Meth.\ A 488 (2002) 493,
10.1016/S0168-9002(02)00515-6. 

\bibitem{totem_Bozzo}
M.~Bozzo, M.~Oriunno, L.~Ropelewski, F.~Sauli, R.~Orava, J.~Ojala, K.~Kurvinen, R.~Lauhakangas, J.~Heino, W.~Snoeys, F.~Ferro, M.~Van Stenis and E.~David, \emph{Design and construction of the triple GEM detector for TOTEM}, IEEE Symposium Conference Record Nuclear Science 2004., Rome, pp.447 - 450 Vol. 1. (2004),
10.1109/NSSMIC.2004.1462231. 

\bibitem{totem_Bagliesi} 
M.~G.~Bagliesi, M.~Berretti, E.~Brucken, {\it et al.}, \emph{The TOTEM T2 telescope based on triple-GEM chambers}, Nucl.\ Instrum.\ Meth.\ A 617 (2010) 134

\bibitem{cms_Colaleo}
  A.~Colaleo, A.~Safonov, A.~Sharma and M.~Tytgat,
  \emph{CMS Technical Design Report for the Muon Endcap GEM Upgrade},
  CERN-LHCC-2015-012, CMS-TDR-013.

\bibitem{alice_Lippmann}
C.~Lippmann, \emph{A continuous read-out TPC for the ALICE upgrade}, Nucl.\ Instrum.\ Meth.\ A 824 (2016) 543,
 10.1016/j.nima.2015.12.007. 

\bibitem{alice_Ketzer}
B.~Ketzer, \emph{A time projection chamber for high-rate experiments: Towards an upgrade of the ALICE TPC}, Nucl.\ Instrum.\ Meth.\ A 732 (2013) 237,

\bibitem{sphenix_Aidala}
  C.~Aidala {\it et al.},
  \emph{sPHENIX: An Upgrade Concept from the PHENIX Collaboration}
  arXiv:1207.6378 [nucl-ex].

\bibitem{Bressan} 
A.~Bressan, M.~Hoch, P.~Pagano {\it et al.}, \emph{High rate behavior and discharge limits in micropattern detectors}, Nucl.\ Instrum.\ Meth.\ A 424 (1999) 321 

\bibitem{Mathis_GEMpaper}
P.~Gasik, A.~Mathis, L.~Fabbietti and J.~Margutti, \emph{Charge density as a driving factor of discharge formation in GEM-based detectors}, Nucl.\ Instrum.\ Meth.\ A 870 (2017), 
10.1016/j.nima.2017.07.042. 

\bibitem{Bachmann} S.~Bachmann, A.~Bressan, M.~Cape\'{a}ns, M.~Deutel, S.~Kappler, B.~Ketzer, A.~Polouektov, L.~Ropelewski, F.~Sauli, E.~Schulte, L.~Shekhtman and A.~Sokolov, \emph{Discharge studies and prevention in the gas electron multiplier (GEM)}, Nucl.\ Instrum.\ Meth.\ A 479 (2002) 294.
 doi:10.1016/S0168-9002(01)00931-7
 
 

 \bibitem{Peskov}
 V.~Peskov and P.~Fonte, \emph{Research on discharges in micropattern and small gap gaseous detectors}, arXiv:0911.0463 (2009)
  
\bibitem{Deisting:2019qda}
  A.~Deisting, C.~Garabatos, P.~Gasik {\it et al.},
  \emph{Secondary discharge studies in single- and multi-GEM structures}, Nucl.\ Instrum.\ Meth.\ A 937 (2019) 168.

\bibitem{Utrobicic:2019tss}
  A.~Utrobicic, M.~Kovacic, F.~Erhardt, M.~Jercic, N.~Poljak and M.~Planinic,
  arXiv:1902.10563 [physics.ins-det].
  
\bibitem{Techtra}
TECHNOLOGY TRANSFER AGENCY TECHTRA SP.Z O.O,~\url{http://techtra.pl/}
  
\bibitem{alpha} Eckert \& Ziegler Nuclitec GmbH, http://www.ezag.com, 2017.

\bibitem{MerlinMPGD} J.~Merlin for the CMS Muon group, \emph{Study of discharges and their effects in GEM detectors}, Talk at MPGD 2019 conference, 07.05.2019, \url{https://indico.cern.ch/event/757322/contributions/3396501/attachments/1839468/3015160/JMerlin_MPGD2019_Discharge_Study_V1_23042019.pdf}




\end{thebibliography}
